\documentclass[10pt,a4paper]{article}
\usepackage{graphicx}
\usepackage{subfigure}
\usepackage{amsmath,amsthm}
\usepackage{amssymb,amsfonts}
\usepackage{authblk}

\newtheorem{definition}{\textbf{Definition}}

\title{Multivariate Normality Test with Copula Entropy}
\author{Jian MA\thanks{Email: majian@hitachi.cn}}
\affil{Hitachi China Research Laboratory}

\begin{document}

\maketitle

\begin{abstract}
	\noindent
	In this paper, we proposed a multivariate normality test based on copula entropy. The test statistic is defined as the difference between the copula entropies of unknown distribution and the Gaussian distribution with same covariances. The estimator of the test statistic is presented based on the nonparametric estimator of copula entropy. Two simulation experiments were conducted to compare the proposed test with the five existing ones. Experiment results show the advantage of our test over the others.
\end{abstract}

\section{Introduction}
Gaussian distribution is one of the most fundamental probabilistic distribution functions in probability and statistics. Normality is assumed in many statistical models and methods and therefore normality test is a basic tool in practice. There are abundant of literature on such test, see \cite{Mardia1980,Henze2002} for the reviews on this topic.

We focus on multivariate normality test in this paper. There are many existing tests on multivariate normality that were proposed based on different concepts, such as characteristics function \cite{Baringhaus1988}, moments \cite{Henze2019}, skewness and kurtosis \cite{Mardia1970}, energy distance \cite{Rizzo2016}, entropy \cite{Vasicek1976}, Wasserstein distance \cite{Hallin2021}, etc. The statistics of these tests are defined as the difference of the properties of the unknown distribution and Gaussian distribution. For example, entropy-based test is defined as the difference of the entropies of the unknown distribution and Gaussian distribution based on the fact that among the distributions with same covariance, Gaussian distribution has the maximum entropy.

Copula Entropy (CE) is a kind of Shannon entropy defined with copula function \cite{Ma2011}. It is a multivariate measure of statistical independence with several good properties, such as symmetric, non-positive (0 iff independent), invariant to monotonic transformation, and particularly, equivalent to correlation coefficient in Gaussian cases. It has been proved that CE is equivalent to mutual information in information theory. 

In this paper, we proposed a multivariate normality test with copula entropy. The test statistic is defined as the difference of CE of the unknown distribution and Gaussian distribution. It is well known that given covariance matrix, the CE of Gaussian distributions can be easily derived analytically. Therefore, the test statistic can be derived by simply estimating the CE of the unknown distribution. This can be achieved with the non-parametric estimator of CE proposed in \cite{Ma2011}.

This paper is organized as follows: Section \ref{sec:ce} introduces the basic theory of CE, the test statistic and its estimator is proposed in Section \ref{sec:test}, simulation experiments will be presented in Section \ref{sec:sim}, Section \ref{sec:con} concludes the paper.

\section{Copula Entropy}
\label{sec:ce}
Copula theory is a probabilistic theory on representation of multivariate dependence \cite{nelsen2007,joe2014}. According to Sklar's theorem \cite{sklar1959}, any multivariate density function can be represented as a product of its marginals and copula density function (cdf) which represents dependence structure among random variables. 

With copula theory, Ma and Sun \cite{Ma2011} defined a new mathematical concept, named Copula Entropy, as follows:
\begin{definition}[Copula Entropy]
	Let $\mathbf{X}$ be random variables with marginals $\mathbf{u}$ and copula density function $c$. The CE of $\mathbf{X}$ is defined as
	\begin{equation}
	H_c(\mathbf{x})=-\int_{\mathbf{u}}{c(\mathbf{u})\log c(\mathbf{u})d\mathbf{u}}.
	\label{eq:ce}
	\end{equation}	
\end{definition}

A non-parametric estimator of CE was also proposed in \cite{Ma2011}, which composed of two simple steps:
\begin{enumerate}
	\item estimating empirical copula density function;
	\item estimating the entropy of the estimated empirical copula density.
\end{enumerate}
The empirical copula density in the first step can be easily derived with rank statistic. With the estimated empirical copula density, the second step is essentially a problem of entropy estimation, which can be tackled with the KSG estimation method \cite{Kraskov2004}. In this way, a non-parametric method for estimating CE was proposed \cite{Ma2011}.

\section{Test on Multivariate Normality}
\label{sec:test}
Given random vector $\mathbf{X}\in R^d$, the hypothesis for multivariate normality test is
\begin{equation}
	H_0: p(\mathbf{x}) \in N_d(,V_x),
\end{equation}
where $N_d$ denotes normal distribution, and $V_x$ is the covariances of $\mathbf{X}$.

The test statistic for $H_0$ is defined as follows:
\begin{equation}
	T_{ce}=H_c(\mathbf{x})-H_c(\mathbf{x}_n),
	\label{eq:tce}
\end{equation}
where $\mathbf{x}_n$ is the Gaussian random vector with the same covariances. The first term in \eqref{eq:tce} can be estimated with the nonparametric CE estimator. The second term in \eqref{eq:tce} can be derived analytically as
\begin{equation}
	H_c(\mathbf{x}_n)=\frac{1}{2}{\log\left(|V_x|\right)}.
\end{equation}

It is easy to know $T_{ce}=0$ for normal distributions.

\section{Simulations}
\label{sec:sim}
We compared our test with several existing tests, including the tests proposed by Mardia \cite{Mardia1970}, Royston \cite{Royston1983}, Henze and Zirkler \cite{Henze1990}, Doornik and Hansen \cite{Doornik2008}, and the energy distance based test by Rizzo and Sz\'ekely \cite{Rizzo2016}.

Two simulation experiments were conducted with bivariate copula function associated with marginals. In the first simulation, the data were generated from a bivariate normal copula with two marginals: one is normal distribution, and the other is exponential distribution. The parameter of bivariate normal copula is 0.8 and the mean and standard deviation of the normal marginal are 0 and 2 respectively. The rate of the exponential distribution is set from 1 to 10. In the second simulation, the data were generated from a bivariate Gumbel copula with two normal marginals. The mean and standard deviation of the two normal marginals are 0 and 2 respectively. The parameter $\alpha$ of the bivariate Gumbel copula is set from 1 to 10. These two simulations generate different non-normality: one by non-normal marginal and the other by non-normal copula. The sample size is 800 for all the simulations. Each simulation was run for 10 times to derive the average of the test statistics.

The \textsf{R} package \texttt{copula} was used for bivariate normal and Gumbel copula. The \textsf{R} package \texttt{copent} was used for estimating CE in the experiments. The \textsf{R} package \texttt{MVN} was used for the implementations of the other tests.

\begin{figure}
	\centering
	\subfigure[CE]{\includegraphics[width=0.48\textwidth]{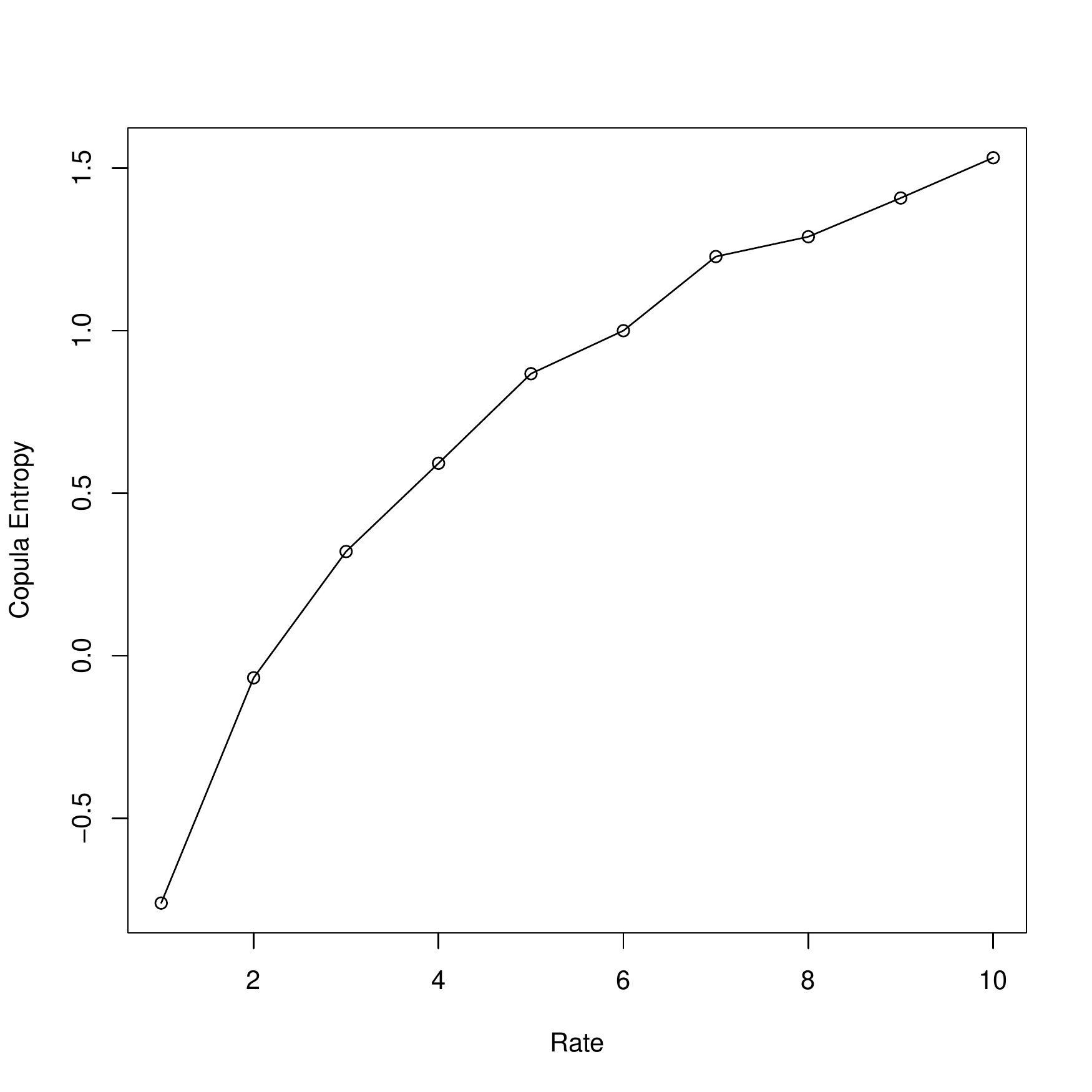}\label{fig:ce1}}
	\subfigure[Mardia]{\includegraphics[width=0.48\textwidth]{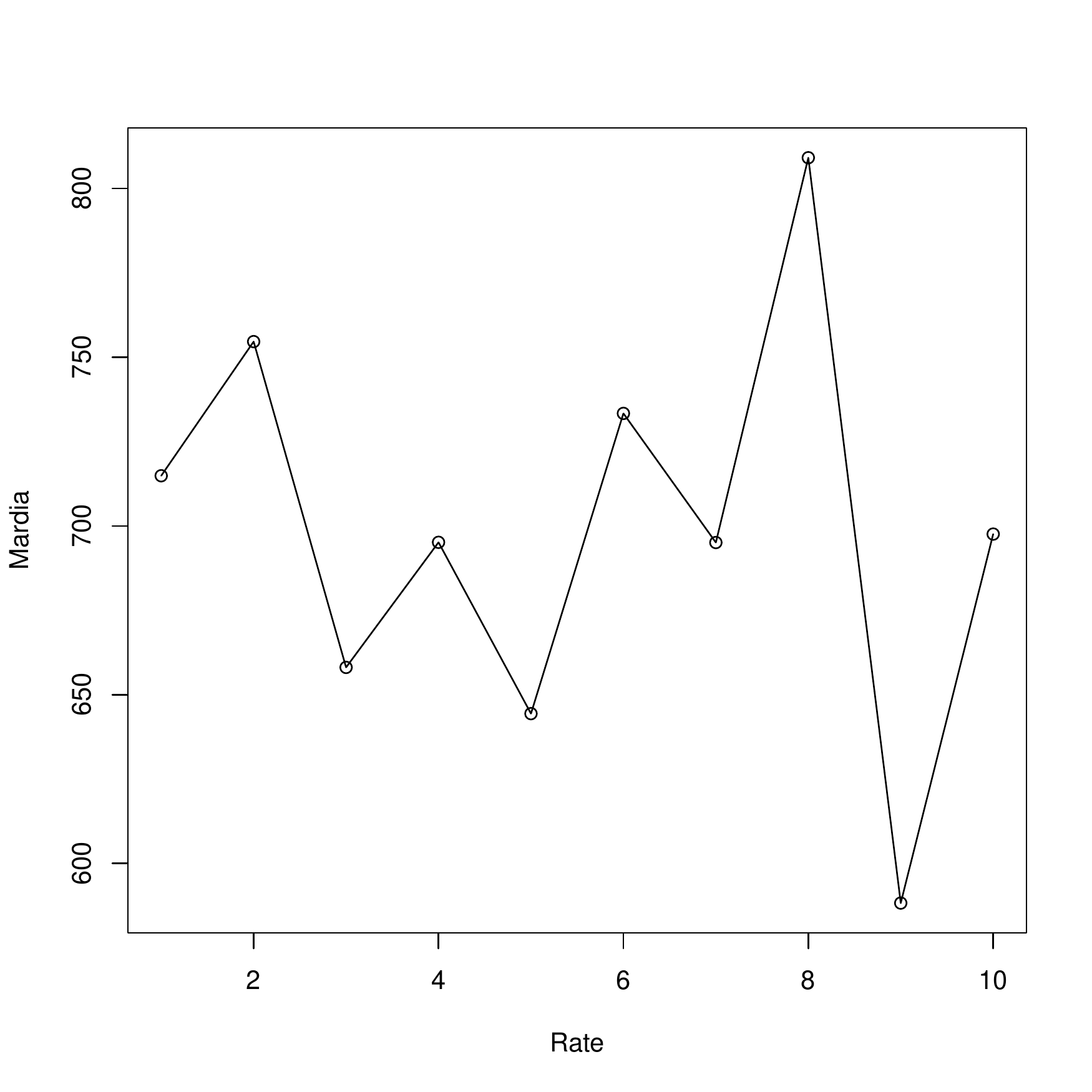}\label{fig:mardia1}}
	\subfigure[Royston]{\includegraphics[width=0.48\textwidth]{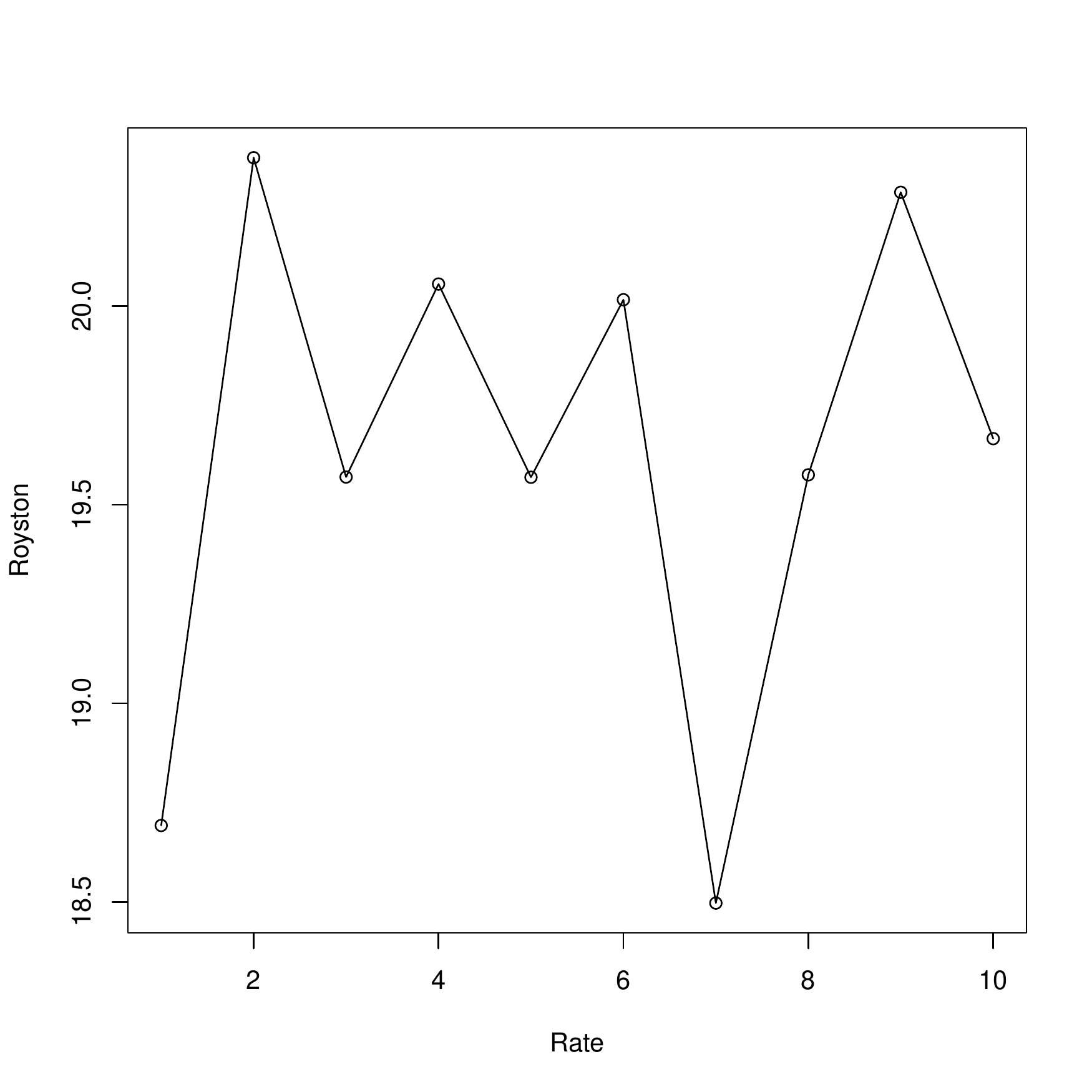}\label{fig:royston1}}
	\subfigure[HZ]{\includegraphics[width=0.48\textwidth]{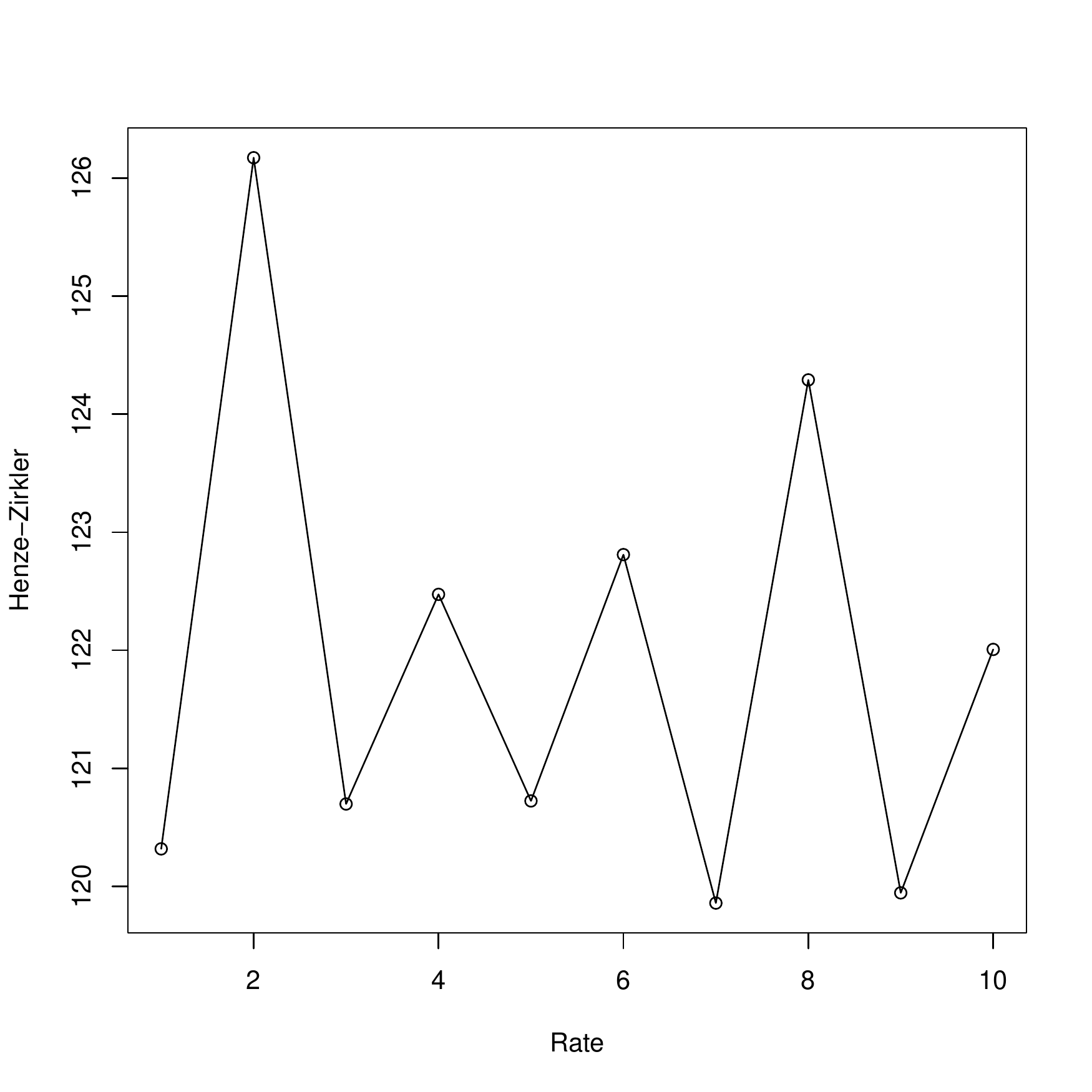}\label{fig:hz1}}
	\subfigure[DH]{\includegraphics[width=0.48\textwidth]{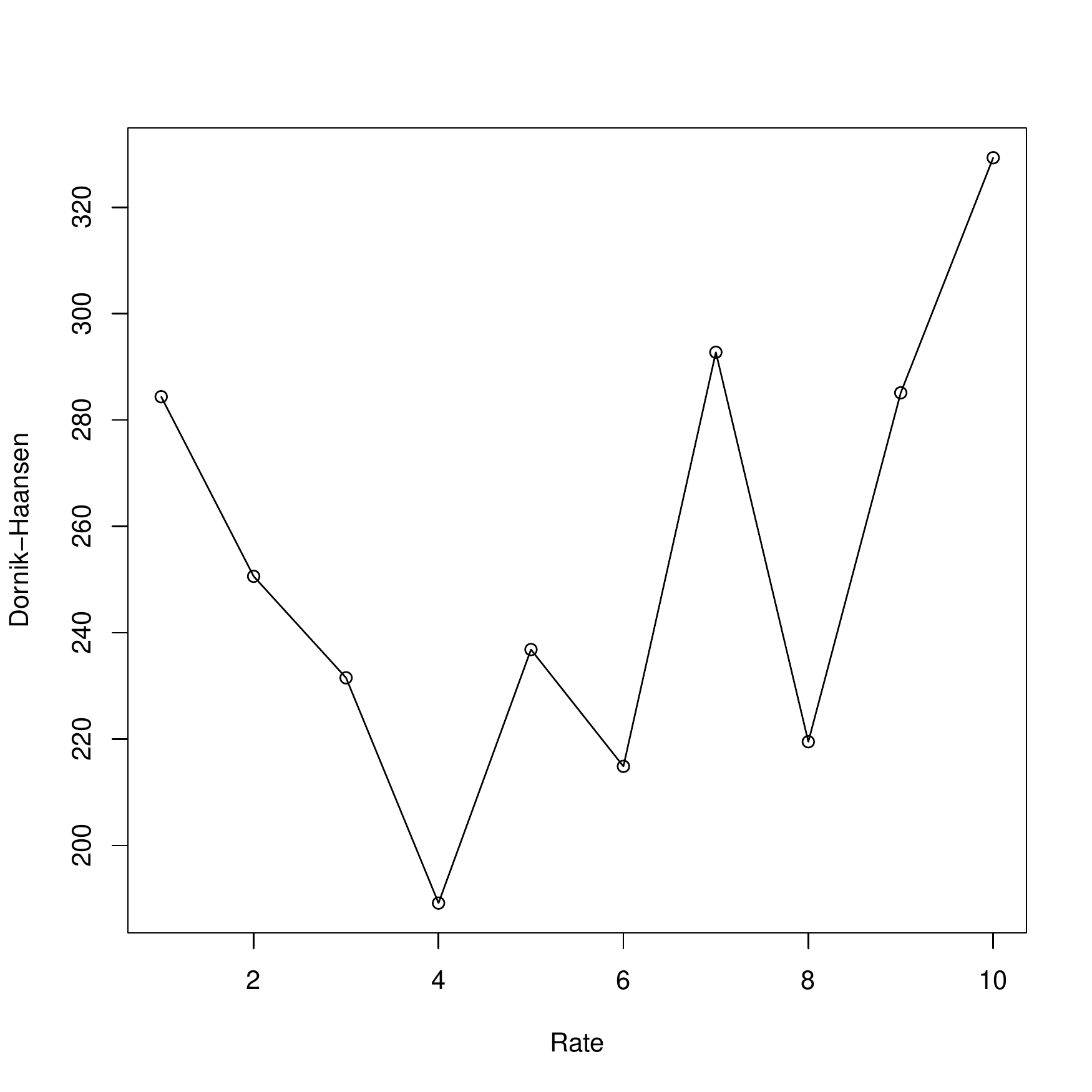}\label{fig:dh1}}
	\subfigure[Energy]{\includegraphics[width=0.48\textwidth]{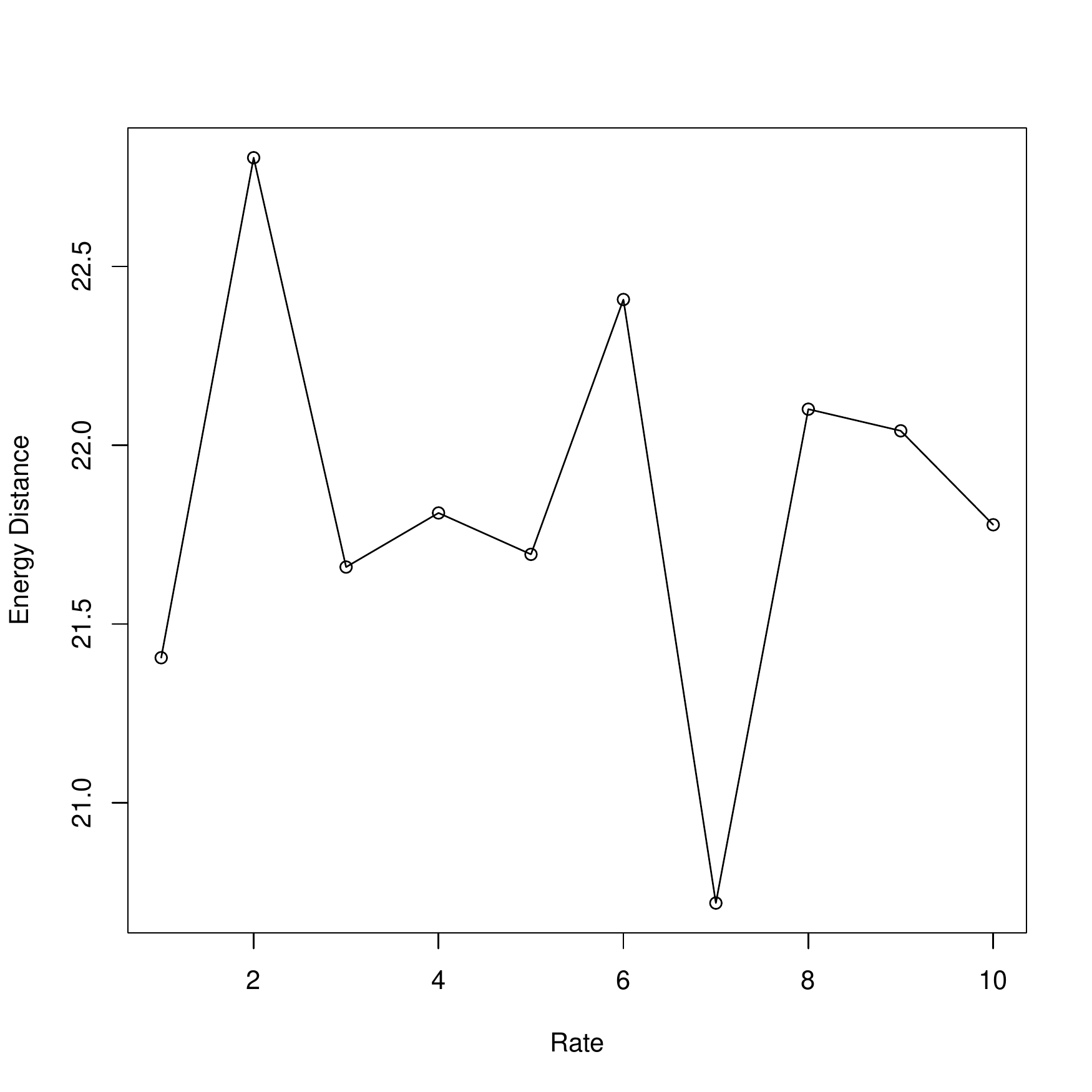}\label{fig:energy1}}
	\caption{Results of the simulation experiments 1.}
	\label{fig:sim1}
\end{figure}

\begin{figure}
	\centering
	\subfigure[CE]{\includegraphics[width=0.48\textwidth]{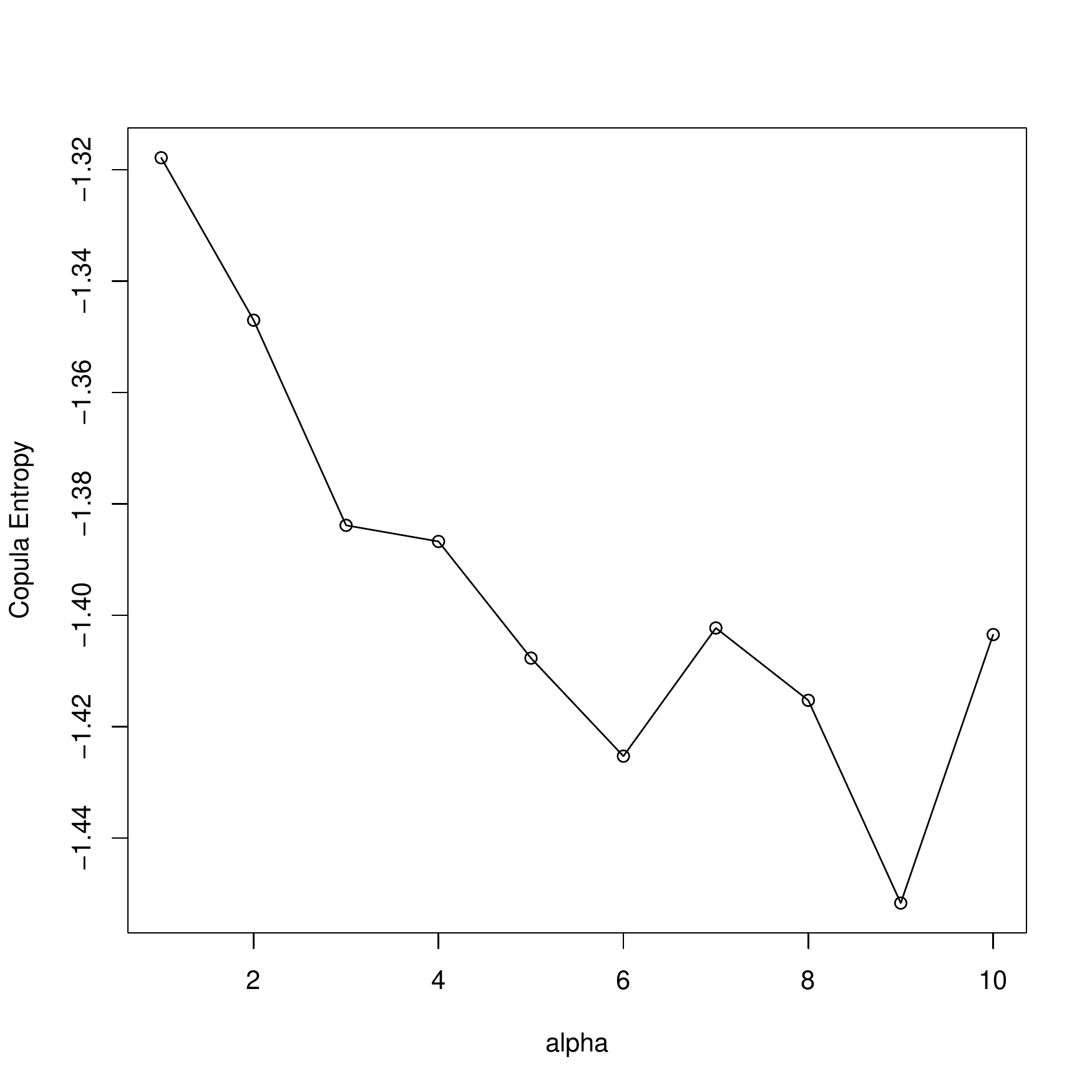}\label{fig:ce2}}
	\subfigure[Mardia]{\includegraphics[width=0.48\textwidth]{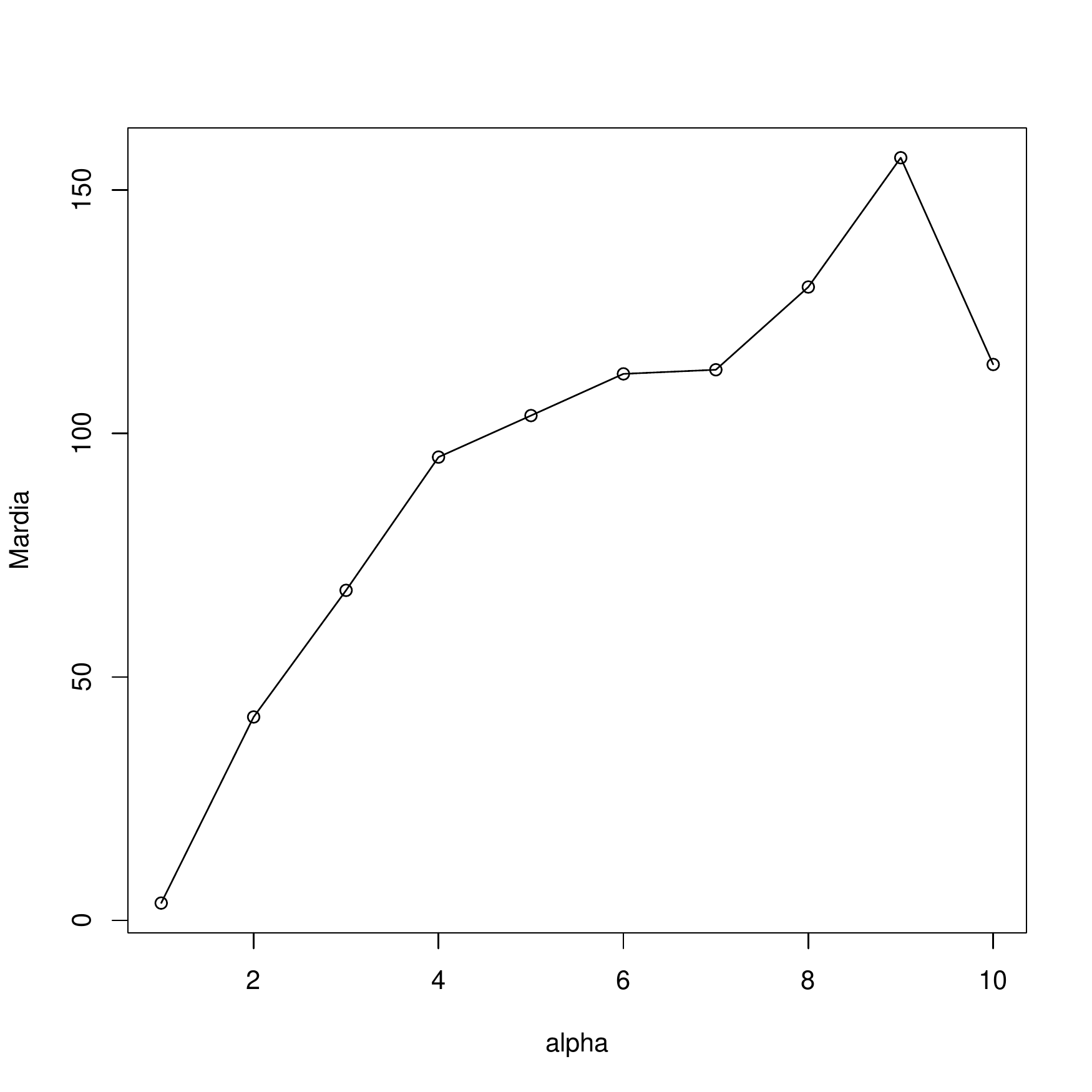}\label{fig:mardia2}}
	\subfigure[Royston]{\includegraphics[width=0.48\textwidth]{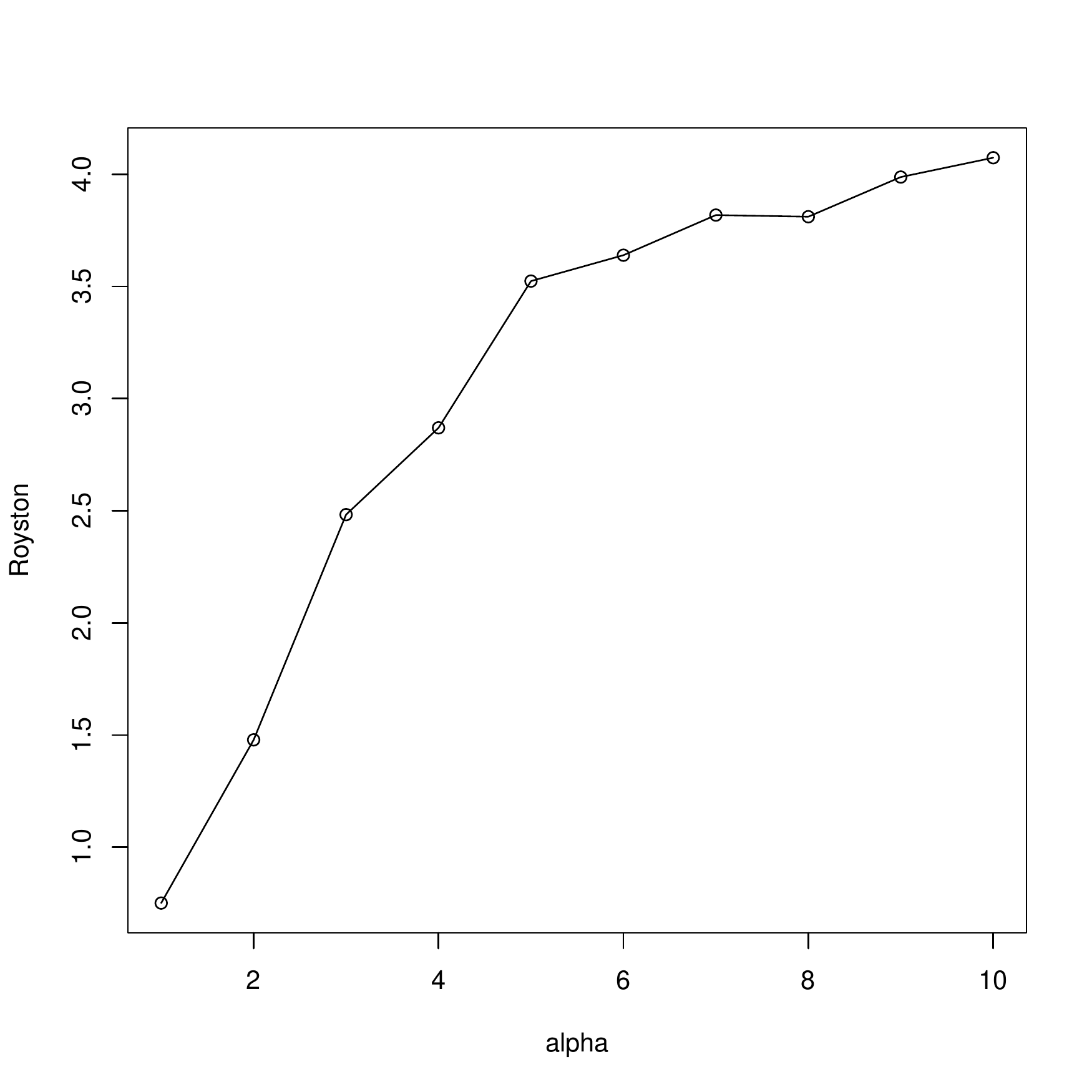}\label{fig:royston2}}
	\subfigure[HZ]{\includegraphics[width=0.48\textwidth]{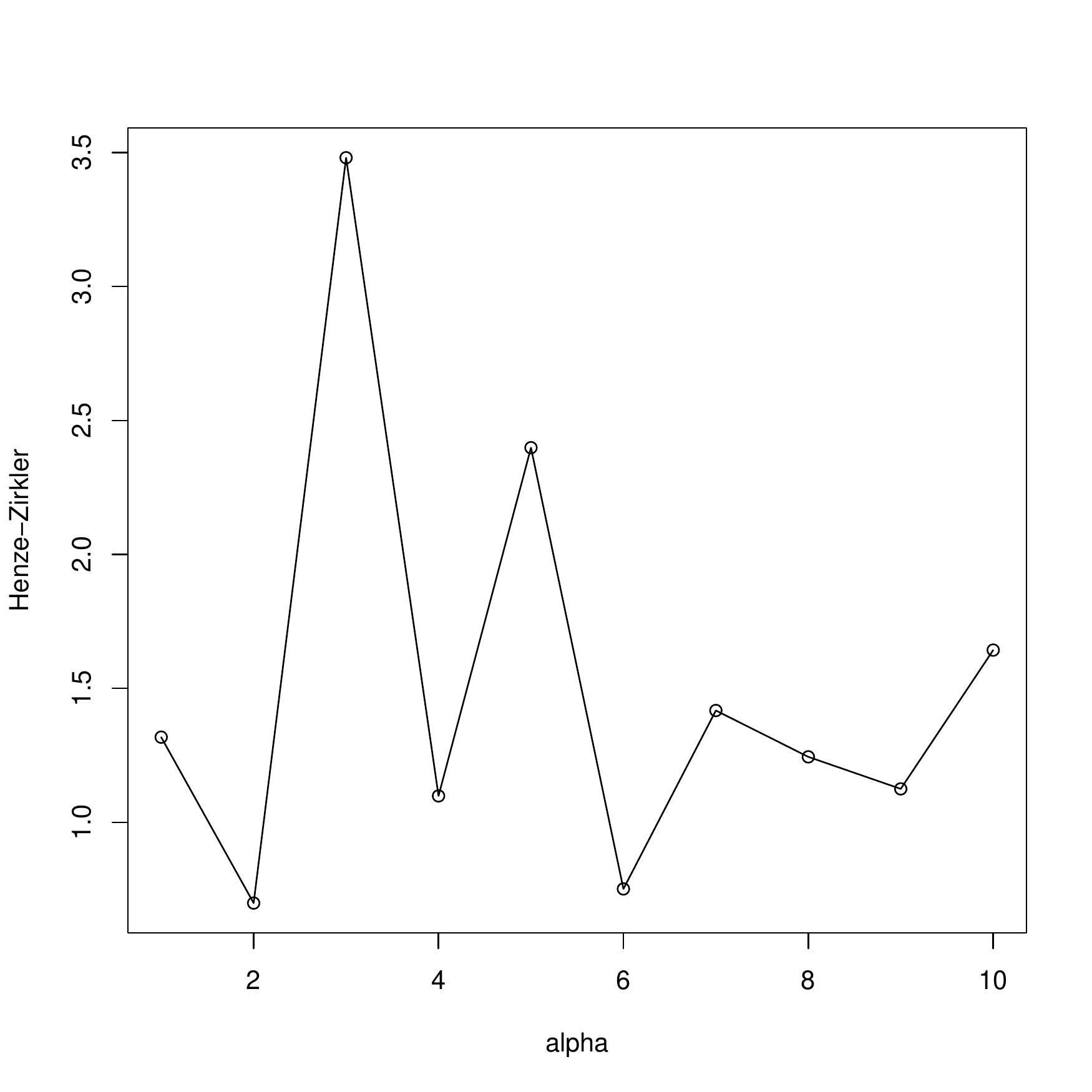}\label{fig:hz2}}
	\subfigure[DH]{\includegraphics[width=0.48\textwidth]{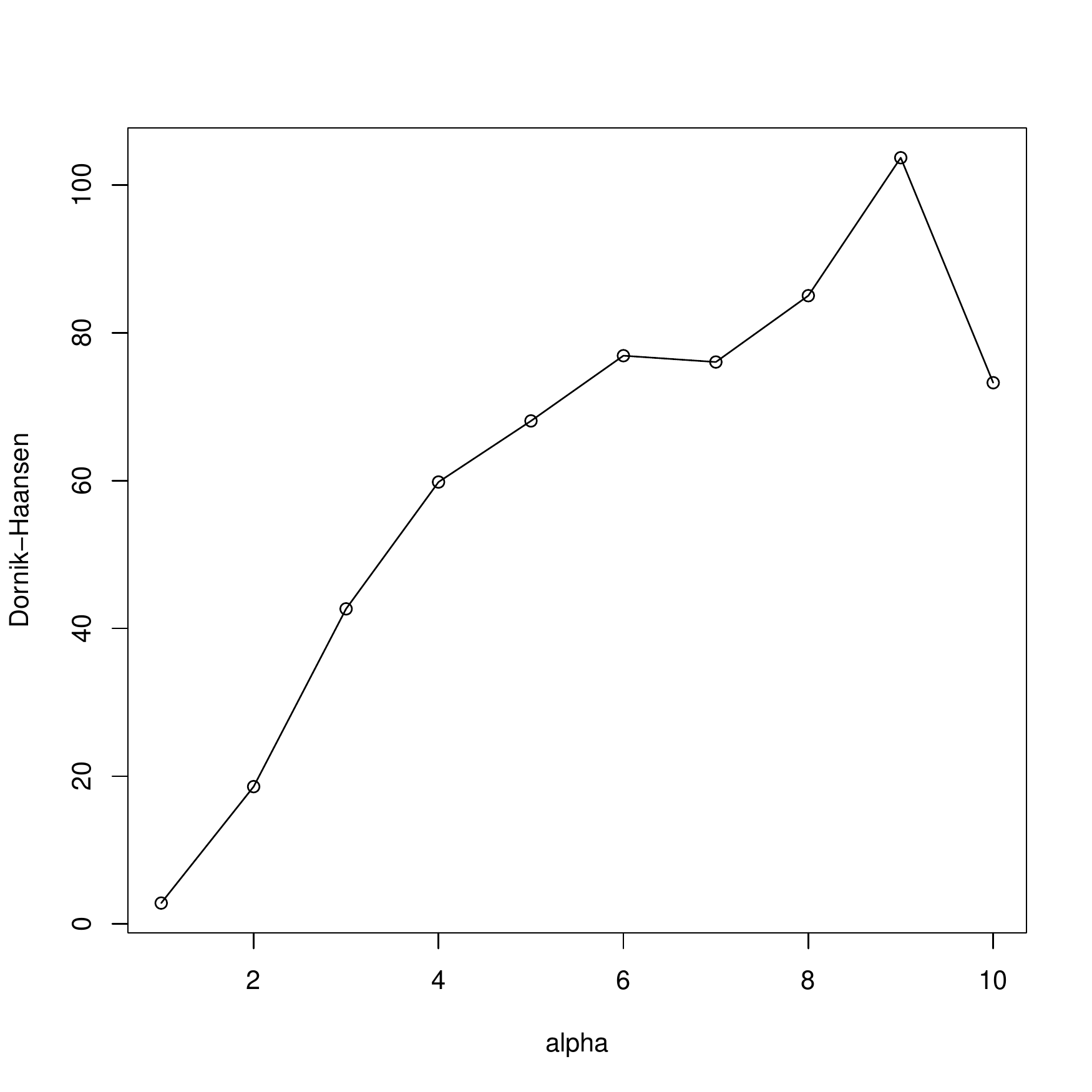}\label{fig:dh2}}
	\subfigure[Energy]{\includegraphics[width=0.48\textwidth]{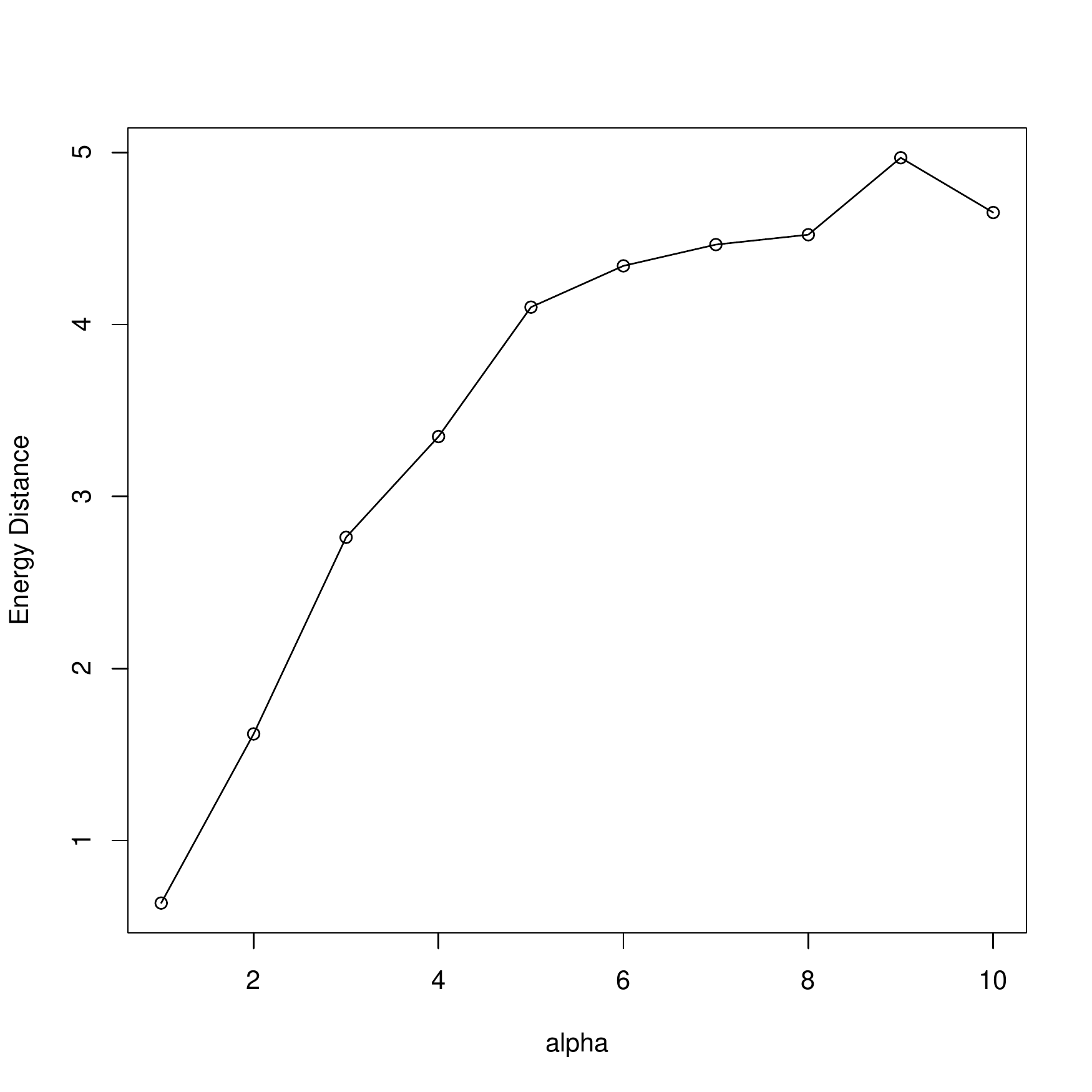}\label{fig:energy2}}
	\caption{Results of the simulation experiments 2.}
	\label{fig:sim2}
\end{figure}

The results of the two simulation experiments are shown in Figure \ref{fig:sim1} and Figure \ref{fig:sim2}. It can be learnt from them that the estimated CE based test statistic reflects the right monotonicity of normality in both experiments. The other five test did not present the monotonicity of normality in the first experiment and four tests (Mardia's, Royston's, DH, and Energy distance) presented the right monotonicity of normality.

\section{Conclusions}
\label{sec:con}
In this paper, we proposed a multivariate normality test based on copula entropy. The test statistic is defined as the difference between the copula entropies of unknown distribution and the Gaussian distribution with same covariances. The estimator of the test statistic is presented based on the nonparametric estimator of copula entropy. Two simulation experiments were conducted to compare the proposed test with the five existing ones. Experiment results show the advantage of our test over the others.

\bibliographystyle{unsrt}
\bibliography{mvnt}

\end{document}